\documentstyle[12pt,epsfig]{article}
\date{}
\title{On measurements of the energy and polarization distributions
of  high energy $\gamma$-beams} 
\author{V.A.Maisheev \thanks{E-mail maisheev@mx.ihep.su} \\
{\it Institute for High Energy Physics, 142284, Protvino, Russia }}

\begin{document}
\large
\maketitle
\def\arcctg{\mathop{\rm arcctg}\nolimits} 
\def\ch{\mathop{\rm ch}\nolimits}
\def\sh{\mathop{\rm sh}\nolimits} 
\normalsize
\begin{abstract}
 A  possibility to measure the energy and polarization distributions
of high intensity $\gamma$-beams is considered. This possibility
is based on measurements of the number of $e^\pm$-pairs in such media,
as laser waves and single crystals. The method may be useful
for  future $\gamma \gamma$ and $e \gamma$-colliders. 
\end{abstract}
\section{ Introduction}
   High energy $\gamma$-beams using in physical experiments
have rather broad energy spectra. The measurement of the energy spectrum
is an important  and primary problem in such experiments. 
Often the intensity of $\gamma$-beams
may be very high, and because of this it is impossible to use 
the number of $\gamma$-quanta counting techniques. 
An excellent example of this
situation is the $\gamma$-beams on  future $e\gamma$ and 
$\gamma \gamma$-colliders\cite{GKS,GKS1,ABE,TES,KEK}. 
In these cases for the correct 
interpretation of  experimental data 
it will be important to know the
spectral luminosity and its polarization characteristics.

In this paper we discuss a possibility to measure the spectral and 
polarization distributions of high intensity $\gamma$-beams  
with   energies of hundreds GeV.
  
\section{Main principles of measurement}

Let us consider a parallel beam of high energy $\gamma$-quanta with
some (unknown) energy  distribution. We assume  that this
distribution is not changed for the time of measurements.
In particular, this situation will take  place on  future 
$\gamma \gamma$-colliders. Really, these $\gamma\gamma$-colliders are 
 accelerators with the periodic action when parameters of beams 
are maintained the same in each cycle. 

It is known that
$\gamma$-quanta are detected by their interactions
with various media.  The process of $e^{\pm}$-pair 
production followed by the development of electromagnetic cascade is the 
main result of such interaction.
In this paper we consider a possibility to measure 
the energy and polarization 
distributions of  high energy $\gamma$-beams with the use of thin layer 
of some specific medium, in which the $e^{\pm}$-pair 
production cross section
is known with a high precision.  

It is well known that the cross sections of $e^{\pm}$-pair production for
 high energy $\gamma$-quanta in amorphous media 
are practically independent 
of their energy \cite{BLP}. However, there exist media, 
in which the cross sections of
$e^{\pm}$-pair production are dependent on the energy and polarization of
high energy $\gamma$-quanta. The monochromatic laser wave \cite{NR}
 and oriented
single crystals \cite{U,Ter} are examples of such media (see Fig.1). 

First we consider the possibility
of measurements of the energy spectrum with the use of 
monochromatic laser wave.  
In this case the cross section of the $e^{\pm}$-pair production has
the following form \cite{KS, MV, MV1}:
\begin{equation}
\sigma_{\gamma \gamma}(z) = \sigma_0(z) + \sigma_c(z) P_2 \xi_2 + 
\sigma_l(z)(\xi_1 P_1 + \xi_3 P_3),
\label{1}
\end{equation}
where $z= {m^2c^4 \over E_\gamma E_l}$ 
is the invariant variable, $E_\gamma$ 
and $E_\l$ are the energies of $\gamma$-quantum and laser photon, $m$ is
the electron mass, c is the speed of light, $\xi_i$ and $P_i, \, (i=1-3)$ 
are the Stokes parameters
 of $\gamma$-quantum and laser photon, respectively.
Functions $\sigma_0,\, \sigma_c,\, \sigma_l$ 
one can find in \cite{KS, MV, MV1}. 
These three functions are different from zero at $ 0< z \le 1$. It 
means that the pair production  in  field of the laser wave is a threshold
process. Note that the variable $z$ is written for the counter motion 
of $\gamma$-quanta and laser wave. Generally, $z= 4m^2c^4/s$, 
where $s$ is the total energy
of the $\gamma$-quantum and the laser photon in their center-of-mass system.

 In the general case of two colliding beams with corresponding densities 
 (numbers of particles per volume unite) $n_1$ and $n_2$,  and
  velocities $\bf{v_1, \, v_2}$ the number of interactions in volume $dV$
 for a time dt is equal to \cite{LL}:
 \begin{equation}
d\nu= \sigma \sqrt{(\bf{v_1}-\bf{v_2})^2-[\bf{v_1},\bf{v_2}]^2}n_1n_2 dV dt,
\label{2}
\end{equation} 
where $\sigma$ is the cross section of interaction. 
In the case when $\bf{v_1}$ and $\bf{v_2}$ 
are directed along the straight line this
equation has the following form:
\begin{equation}
d\nu= \sigma |v_1-v_2|n_1 n_2 dV dt
\label{3}
\end{equation}
For $\gamma \gamma$-interactions  these equations are simplified:
\begin{equation}
d\nu= (1- \cos{\theta}) \sigma  n_1 n_2 dV dt,
\label{4}
\end{equation}
\begin{equation}
d\nu = 2 \sigma n_1 n_2 dV dt
\label{5}
\end{equation}
where $\theta$ is the angle between directions of motion of these beams.

Let $n_\gamma$ and $n_l$ are the densities of 
high energy $\gamma$-beam
and laser beam, correspondingly. 
In this case it would appear reasonable that
$n_l >> n_\gamma$. It means that we can neglect the variations 
of  $n_l$-value in the considered process. Now we can find 
the total number of interactions in the thin volume $V$ of laser bunch
\begin{equation}
n_i(E_l,P_1,P_2,P_3)  = 2 n_l \delta x 
 \int_{E_{\gamma, min}}^{E_{\gamma, max}} 
{N_\gamma(E_\gamma) \sigma_{\gamma \gamma} (E_\gamma, E_{l}) dE_\gamma },
\label{6}
\end{equation}
where $N_\gamma$ is the energy distribution of $\gamma$-beam, 
$E_{\gamma,min}$ and $E_{\gamma,max}$ are the minimum and maximum 
energies of $\gamma$-quanta, $\delta x$  
is the thickness of the laser bunch
($\delta x n_l \sigma_{\gamma \gamma} <<1 $).
It can be assumed
that the  photon energy in Eq. ({6}) is variable. It means that 
 the laser frequency is changed
for each or some cycles of  $\gamma \gamma$-collider. Besides, we
propose that the number $n_i$ can be measured 
with the help of some detector  
and it can be a scintillation counter. The number of photons 
creating from passing relativistic particles
through the counter in a time of some picoseconds are
proportional to the number of these particles. It may be  another
type of
detector, for example, a current transformer, etc.

However, the cross section $\sigma_{\gamma \gamma}$ depends on 
the Stokes parameters
of the $\gamma$-beam and laser wave. Because of this, one can duplicate
the total number of measurements of $n_i$-value. 
One-half of measurements 
must be done with the circular polarization 
of laser wave $P_2 = 1$ and 
the remaining part - with the $ P_2=-1$. Thus we can 
get the following equations:
\begin{eqnarray} 
n_0(E_l) = (n_i(E_l,0,1,0) + n_i(E_l,0,-1,0))/2= \\ \nonumber 
= 2n_l\delta x  \int_{E_{\gamma, min}}^{E_{\gamma, max}}
{N_\gamma(E_\gamma) \sigma_{0} (E_\gamma, E_{l}) dE_\gamma },
\label{7}
\end{eqnarray}
\begin{eqnarray} 
n_c(E_l) = (n_i(E_l,0,1,0) - n_i(E_l,0,-1,0))/2= \\ \nonumber
= 2n_l\delta x  \int_{E_{\gamma, min}}^{E_{\gamma, max}}
{N_\gamma(E_\gamma) <\xi_2(E_\gamma)>
 \sigma_{c} (E_\gamma, E_{l}) dE_\gamma },
\label{8}
\end{eqnarray}
where $<\xi_2(E_\gamma)>$ is the circular polarization
everaged over set of $\gamma$-quanta with
the energy equal to $E_\gamma$ . Eqs. ({7})-({8})  represent
the system of two linear integral equations relative to functions
$N_\gamma(E_\gamma)$ and $N_\gamma(E_\gamma) <\xi_2(E_\gamma)>$.
From these equations one can find the energy and 
circular polarization distributions 
of $\gamma$-beam, if the dependencies $n_0$ and $n_c$ are known.  
Obviously, similar  equations may be written 
for $<\xi_1>$ and $<\xi_3>$ Stokes parameters of $\gamma$-beam. 
It should be noted that Eqs. ({4})-({5}) are valid 
 when the intensity of laser wave is not very high 
 (see \cite{KS, MV}).        

 The pair production process in single crystals \cite{U,Ter}
 may also be  used for the
determination of the energy spectrum and 
linear polarization of $\gamma$-beams.
Generally, this process  depends on the two angles of orientation
relative to direction of $\gamma$-quanta propagation.
 When $\gamma$-beam move near strong
crystallographic plane  and far from axes, the process is described
 with the help of the one orientation angle
$\theta_c$
with respect to the plane. Fig.1c illustrates the pair production cross
sections for unpolarized $\gamma$-quanta propagating in a silicon 
single crystal near (110) plane. One can see that the cross sections 
are changed very strongly at variations of the orientation angle 
in the range 0.2 - 1 mrad. Unlikely, the cross section of pair
production in single crystals depends on the linear polarization and is
independent of the circular polarization \cite{U,Ter} 
(nevertheless, see \cite{MV3}). 
It means that only $<\xi_1>$
and $<\xi_3>$ Stokes parameters  may be determined with the help of 
the single crystals.
 
However, the circular polarization of $\gamma$-beam can be transformed
into a linear one with the help of laser bunch \cite{KS,MV,MV2},
 and thereafter can be analyzed by a single crystal. 
Note that this transformation
take a place without intensity losses of $\gamma$-beam 
(if $E_{\gamma,max} < m^2c^4/E_l $).        

 Eq. ({6})  can be rewritten in the following differential form:
\begin{equation}
{dn_i\over dE_e}(E_e,E_p)  = n_2 \delta x
\int_{E_{\gamma,min}}^{E_{\gamma,max}} 
 N_\gamma(E_\gamma) {{d\sigma_{\gamma \gamma} \over dE_e}
(E_\gamma,E_e, E_p) dE_\gamma} ,
\label{9}
\end{equation}
where $ d\sigma {\gamma \gamma} \over dE_e$ is the differential cross
section for production $e^\pm$ pair when one of the particles 
has energy equal to $E_e$, $E_p$ is a possible parameter such
as an orientation angle, laser photon energy, etc.  
This equation is written for the fixed target
$(v_2=0)$.   
In principle, one can measure the ${dn_i / dE_e}$-value with
the use of the magnetic analysis. Then Eq. ({9}) is a linear 
integral equation and its solution is the energy 
distribution of $\gamma$-beam. 
It is obvious,  the measurement of the energy distribution 
of $e^\pm$-pairs
allows one to use  various media, for example, amorphous ones.
Note,  equations similar to Eqs. ({7})-({8}) one can
write in differential form too.

 Real $\gamma$-beams have some angular divergence. The constancy of the
cross section $\sigma_{\gamma \gamma}$ in the limits of the divergence
is necessary for using Eqs. ({6})-({9}).
 One can see
from Eq. ({4}) that this condition are true for the laser medium, if
the divergence less then tens mrad.
In the case of the usage of  silicon single crystals oriented near (110)
plane  the divergence $\delta \phi$ of the $\gamma$-beam must satisfy
(only in one direction) the following condition:
 $\delta \phi / \theta_c \leq 0.02$. 
   
\section{Simulation}

In this section we present some calculations, which can help to
feel the problems of realization of energy and polarization measurements  
on the future $\gamma \gamma$-colliders. For our calculations we select 
$E_{\gamma, max}= 400$ GeV. This energy corresponds to     
the electron beam energy about $500$ GeV. 

For a parallel and pointlike electron beam the number 
of scattered photons
with the energy in the range from $E_\gamma$ to $E_\gamma + dE_\gamma$ 
 \cite{GKS,GKS1} is
\begin{equation}
\tilde{dN_{\gamma}} = N_e {k \over \sigma_{ce}}
{d\sigma_{ce} \over dE_\gamma} dE_\gamma
\label{10}
\end{equation}
where $N_e$ is the number of electrons, $k$ is the conversion coefficient,
$\sigma_{ce}$ is the cross section of the Compton effect. 
For simulations we 
select $\gamma$-beam energy distribution in the form:
$d\sigma_{ce}/dE_\gamma$. The form of the energy and polarization 
distributions depends on the helicities $\lambda,\, P_c$  
of electron and laser 
beams \cite{GKS1}. Fig.2 illustrates these distributions for two cases:
1)$2\lambda=1,\, P_c=-1$ ; 2)$2\lambda =-1,\, P_c=-1 $.   

  We calculate the $n_0$ and $n_c$-values
 (see Eqs. ({7}-{8})) for the above-mentioned
the energy and polarization distributions of $\gamma$-beam. Fig.3
illustrates these calculations in 
the laser wave and silicon single crystal.  
Note that $n_c$ is equal to zero for single crystals. 
One can see that these functions are simple  enough and 
have smooth shapes.    

Knowing  $n_0$ and $n_c$-functions one can make an attempt to find the
energy and polarization distributions of $\gamma$-beam for the
above-mentioned cases. The simplest way to do it is to represent of
the unknown  distributions in the form of polynomials  
\begin{equation}
N_\gamma  = \sum_{k=0}^{N}a_k E_\gamma^k \, ,
\label{11}
\end{equation} 
\begin{equation}
N_\gamma <\xi_2> = \sum_{k=0}^N b_k E_\gamma^k \, ,
\label{12}
\end{equation}
where $a_k$ and $b_k$ are the constant coefficients. 
Substitution of Eqs. ({11})-({12}) in Eqs. ({7})-({8})  
allows to obtain the two systems of 
linear algebraic equations for different  $E_{l}$-values. The  
solutions of these equations are $a_k$ and $b_k$-coefficients.

We select N=7 for calculations . We also select the 8 various 
$E_{l}$-values in the follows manner:
$E_{l}= E_{l,min}+\delta E_{l} k,\,  (k=0-7)$.  Here $E_{l, min}$ and 
 $\delta E_{l}$  are
the minimal energy  and step of energy variation for the laser photon.
Some series of 
calculations were done, in which $E_{l,min},\delta E_{l} $,  are varied 
  from 0.8 and 0.25 eV to 1.5 and 1 eV, correspondingly. 
The obtained in such a manner  approximate polynomials are described
good enough  as the whole  energy 
and polarization distribution for two cases.
Fig.2 illustrates the typical description with the help of polynomials
 when $E_{l,min}= 1$ eV  and $\delta E_{l} = 0.25 eV $. 
One can see that the narrow range near the end of energy and polarization 
distributions (from 390  to 400 GeV)  
are described incorrectly for the second 
case. However, if we select $E_{l,min}=0.7$ eV and $\delta E_{l}= 0.05$ eV,
then the polynomial description in a  range from 300 to 400 GeV
becomes good. Note, 
it is impossible to describe properly
the origin of all  distributions because of the threshold in 
$\gamma$-beam and laser photon interactions (see Fig.1). The agreement
on fig.2 between the  polynomial and true distributions 
at small energies one can explain
 by the simple forms of curves or by  randomness. Similar calculations
were also done  with the use of the cross section 
in silicon single crystals 
(see Fig.2c)
at variations of the orientation angle from 0.2 to 1 mrad. 
These calculations
show good agreement between the computed and initial energy distributions. 

We suppose that there is no problem to solve the integral 
equations  numerically, 
if the $n_0$ and $n_c$-functions are known with a high precision. However,
these functions will be obtained as result of measurements. 
Because of this,
it is important to understand 
the influence on the reconstruction of the
energy and polarization distributions such  
factors as fluctuations and
systematic errors.     
 Our estimations on the base of the polynomial model show 
that solutions of Eqs. ({7})-({8}) are stable enough 
relatively to small uniformly-distributed shifts
of $n_0,n_c$-values (of about 1\% to their absolute values).               
On the other hand, small (about 0.01\%-0.001\%) 
fluctuations of the above pointed
values do not allow one to describe satisfactory 
the energy and polarization
distributions with the help of polynomials at N=7. However, this result is
obvious: fluctuations  distort the true distributions and 
more high degree of polynomials or another class of
approximation functions are required for their description. 
We think that for the discount of fluctuations 
it will be properly to correct 
the measured $n_0$ and $n_c$-values, for example, by the method of 
least squares.

\section{Discussion}

In this paper we have considered the possibility to measure 
the energy and polarization
distributions of high energy $\gamma$-beam with intensity which does not
allow to counter every $\gamma$-quantum. This situation is expected on   
 future $ e\gamma$ and $\gamma \gamma$-colliders.
One of main parameters of these colliders is the luminosity. The 
relations for spectral luminosity  one can find in \cite{GKS, GKS1}. 
In the general
case, the  knowledge of the energy and polarization distributions 
does not determinate
completely the spectral luminosity and such important characteristics 
as the everaged $<\xi_i\xi_j>,\, (i,j =1-3)$ Stokes parameters. 
However, we
think that measurements of the above pointed 
distributions will be very useful
for determination of the luminosity but further investigations in this
field are required. Besides, in  specific cases 
(but important for practice)
the spectral luminosity is factored, and the energy and 
polarization         
distributions of both the $\gamma$-beams
are multipliers in the product of some values. The spectral luminosity of 
$e\gamma$ and $\gamma \gamma$-colliders 
will be determined by the use of the various 
calibration processes \cite{GKS,GKS1}.
Contrary to this practice the solution of the above pointed 
problem allows
one to carry out the measurements for some hundreds cycles.
\section{Conclusion}

A new possibility for the measurements of the energy and
 polarization distributions
of $\gamma$-beams in  hundreds GeV is proposed. This possibility 
may be useful for practical applications on  future $e\gamma$ and 
$\gamma \gamma$-colliders. However, 
further development of this possibility
is required. In particular, it is necessary to find  solutions 
of the following problems:

i) the choice of optimal algorithm for solution of Eqs. ({7})-({9});

ii) the reconstruction of spectral luminosity on the base
of measurements of energy distributions of $\gamma$-beams;

iii) the  choice of the detector type for registration of $e^\pm$-pairs;

iv) the estimation of the background and its minimization.  

At the present time the considered method may be investigated
experimentally on  $\gamma$-beams of electron accelerators \cite{VG}. 
 
 We believe that another important problem may be solved on the base
of our consideration. This is the problem of experimental determination
of the equivalent photon spectrum in peripheral collisions of relativistic
ions \cite{B}.

 The author would like to thank V.G. Vasilchenko for useful discussion.

\newpage
\begin{figure} 
\begin{center}
\parbox[c]{12.5cm}{\epsfig{file=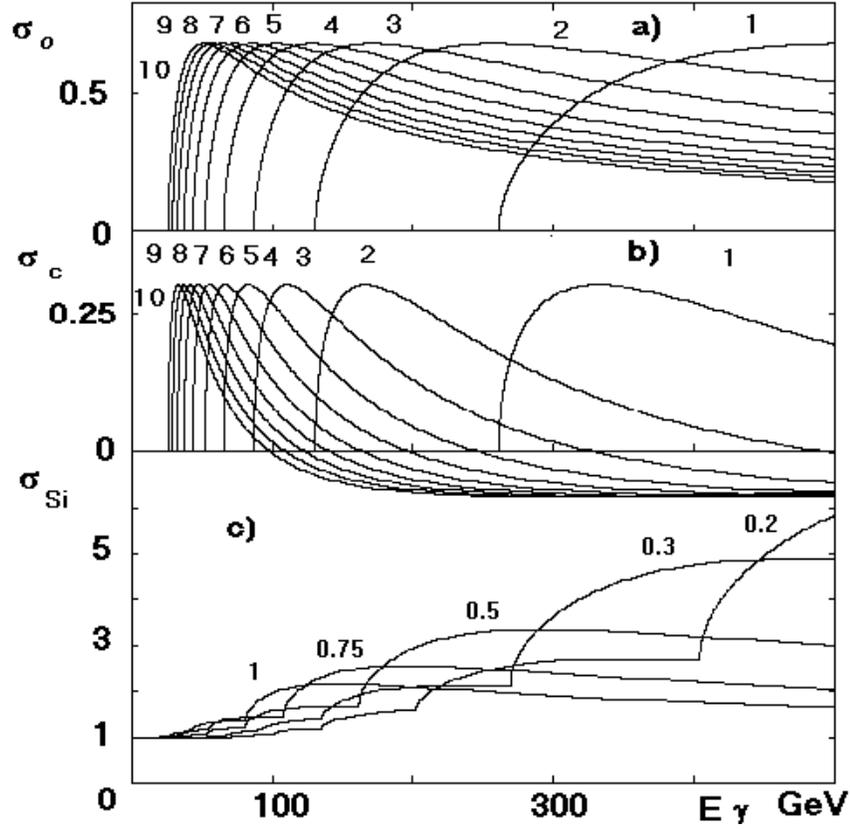,width=12 cm}}
\parbox[c]{15cm}{\caption{
 The cross sections of $e^\pm$-pair production $\sigma_0, \, \sigma_c$
in laser wave (a,b) and $\sigma_{Si}$ in (110) silicon 
single crystal plane (c)
as functions of $\gamma$-quantum energy. The numbers near curves are
the energy of laser photon (a,b) in eV and orientation angle (c) in mrad. 
$\sigma_0$ and $\sigma_c$ is measured in $\pi r_e^2$ units, where $r_e$ is
electron radius. $\sigma_{Si}$ is normalized on the cross section when the
crystal is not aligned.  
	      }}  
\end{center} 
\end{figure}

\begin{figure}
\begin{center}
\parbox[c]{12.5 cm}{\epsfig{file=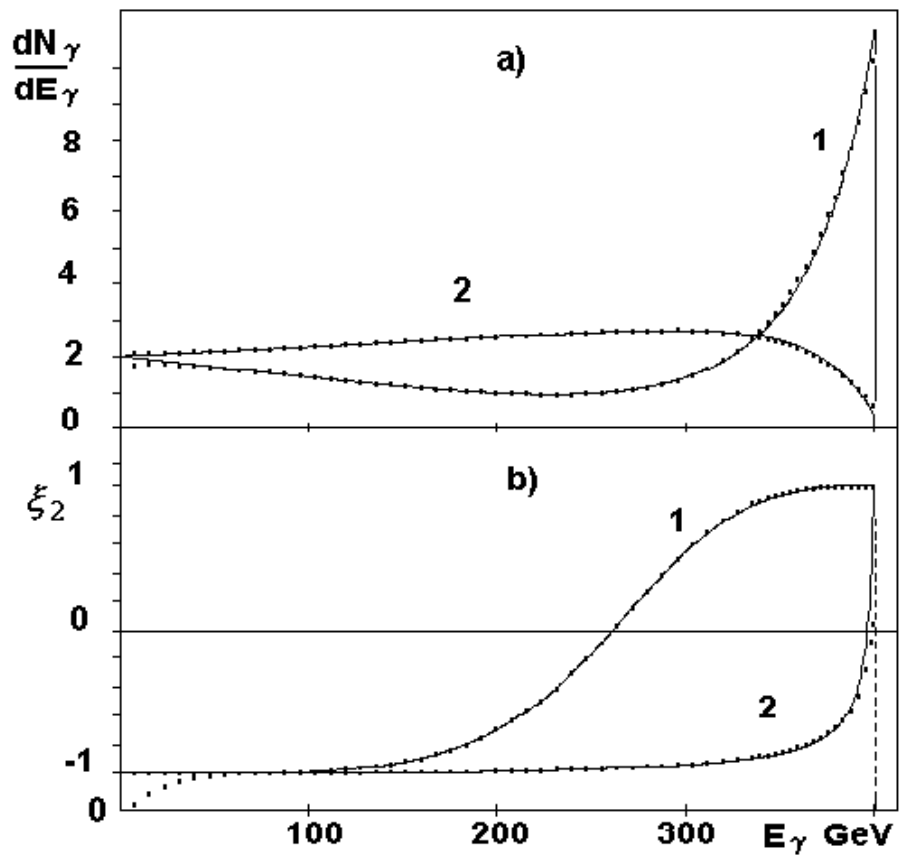, width=12 cm}}
\parbox[c]{15cm}{\caption{
 The energy (a) and circular polarization (b) distributions of 
$\gamma$-beam for two cases: 1) 2$\lambda$=1, $P_c$= -1; 2)2$\lambda$=-1,
$P_c$=-1. The pointed curves are result of reconstruction 
(see text for detail).  
	      }}  
\end{center}
\end{figure}

\begin{figure}
\begin{center}
\parbox[c]{12.5 cm}{\epsfig{file=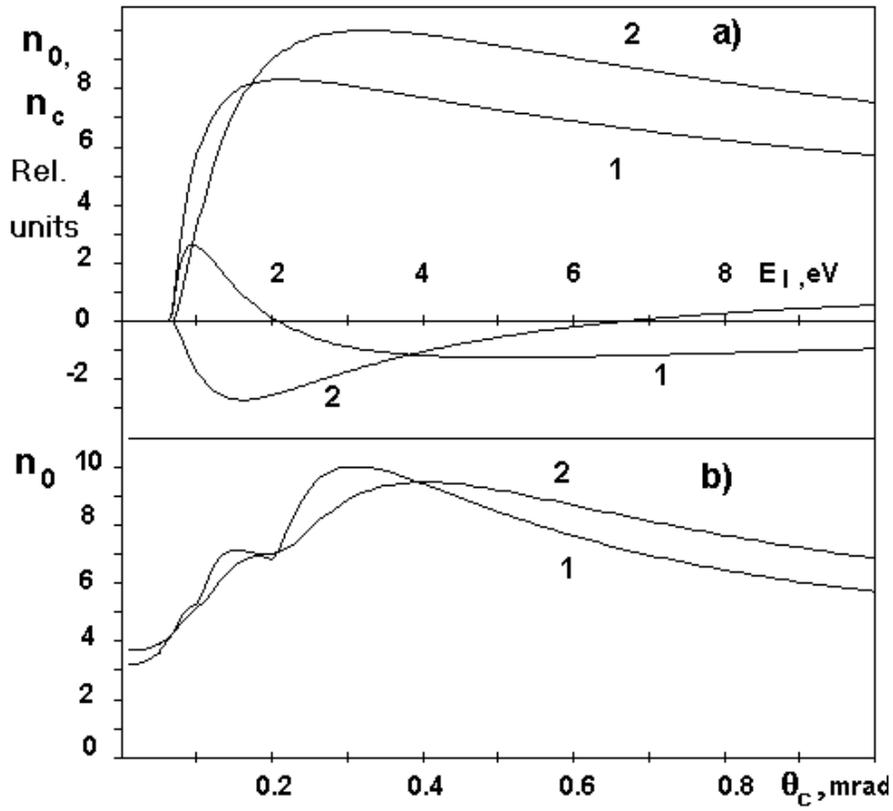, width=12 cm}}
\parbox[c]{15cm}{\caption{
 a)Calculated $n_0$ and $n_c$-values as functions of laser photon energy;
b)calculated $n_0$-values as functions of the orientation angle of silicon
single crystal relative to (110) plane. Numbers near curve correspond to:
1) $2\lambda$=1, $P_c$=-1; 2)2$\lambda$=-1, $P_c$=-1. 
	      }}  
\end{center}
\end{figure}

\end{document}